# Sensitivity of the Blue Loops of Intermediate-Mass Stars to Nuclear Reactions

Ghina M. Halabi[a] and Mounib El Eid[a]

[a]American University of Beirut, Department of Physics, P.O. Box 11-0236, Riad El-Solh, Beirut, Lebanon, e-mail:gfm01@aub.edu.lb

**Abstract.** We investigate the effects of a modification of the $^{14}N(p,\gamma)^{15}O$ reaction rate, as suggested by recent evaluations, on the formation and extension of the blue loops encountered during the evolution of the stars in the mass range $5M_\odot$ to $12M_\odot$. We show that the blue loops of stars in the mass range $5M_\odot$ to $8M_\odot$, that is the range of super ABG stars, are severely affected by a modification of the important $^{14}N(p,\gamma)^{15}O$ reaction rate. We also show that the blue loops can be restored if envelope overshooting is included, which is necessary to explain the observations of the Cepheid stars.

## INTRODUCTION

A remarkable and intriguing feature of the evolution of stars in the mass range $5M_\odot$ to $12M_\odot$ is that they develop blue loops after the onset of core helium burning, when the star reaches the red giant branch (RGB). This crossing of the Hertzsprung-Russell diagram (shortly HRD) towards higher effective temperatures and back to the RGB is commonly called "blue loop" in the literature. It is well established that the blue loops are necessary for explaining the observed non-variable yellow giants, supergiants and the δ-Cepheids in open galactic clusters ([9], [10], [6], [7]).

We focus on the effect of the $^{14}N(p,\gamma)^{15}O$ reaction on the properties of the blue loops in intermediate mass stars (IMS), because it controls the energy production during shell-hydrogen burning which is a primary cause of the loop formation, especially in the mass range $5M_\odot$ to $7M_\odot$. In addition to being a crucial rate that determines the efficiency of the CNO cycle, this particular rate is still not settled on experimental ground and is still subject to significant modification. As we shall see in the present work, the recent suggested modifications of this rate (see [1])) have a strong impact on the efficiency of shell hydrogen burning (hereafter briefly shell H-burning) which, in turn, significantly influences the behavior of the loops in a certain mass range.

The blue loops are also sensitive to mixing processes like overshooting. Envelope overshooting is found to promote the formation and extension of blue loops in IMS.

## Stellar Model Calculations

### Stellar evolution code

The evolutionary sequences presented in this work are obtained using the stellar evolution code described by [4], [5] and references therein. The reaction rates we have used are those recommended by the "JINA REACLIB" database [3]. For the $^{14}N(p,\gamma)^{15}O$ rate (hereafter referred to as N14 rate), we use three compilations: (a) The rate provided by NACRE [2], which we refer to as the N-rate, (b) a new evaluation of this rate, which we call C-rate in the following, is described in detail elsewhere [8] and (c) the rate given by the LUNA collaboration [9], hereafter briefly referred to as the L-rate. We could show that both the C-rate and the L-rate are significantly lower compared to NACRE compilation for the relevant temperature range by approximately a factor of 2.

### Results

We present a set of evolutionary sequences for stars in the mass range $5M_\odot$ to $12M_\odot$ with the initial composition (X,Y,Z)=(0.7,0.28,0.02). These stars are known to develop blue loops after the onset of core helium burning on the red giant branch. Such loops are required to explain the observed Cepheid stars and the yellow giants.

The stellar models are calculated from the zero-age main sequence till the end of core helium burning. The evolutionary tracks in the HRD are shown in Figures 1, 2 and 3 as they are obtained with the same input physics except the N14 rate, for which the three different rate compilations are adopted. The evolutionary tracks show interesting effects of the N14 rate on the blue loops, and a detailed analysis of these results and the physics underlying them is discussed in [8]. However, here we only mention the main findings, which are summarized as follows:

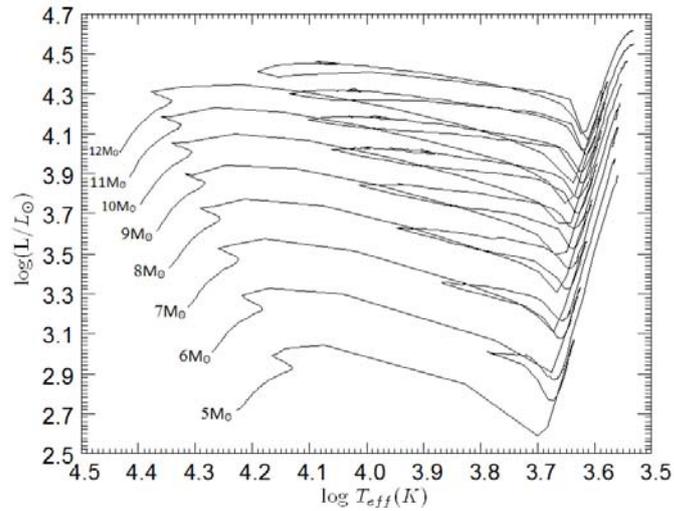

**FIGURE 1.** Evolutionary tracks in the HRD for the indicated stellar masses evolved from the zero-age main sequence till the end of core helium burning. The N-rate is used for the $^{14}N(p,\gamma)^{15}O$ reaction, see text.

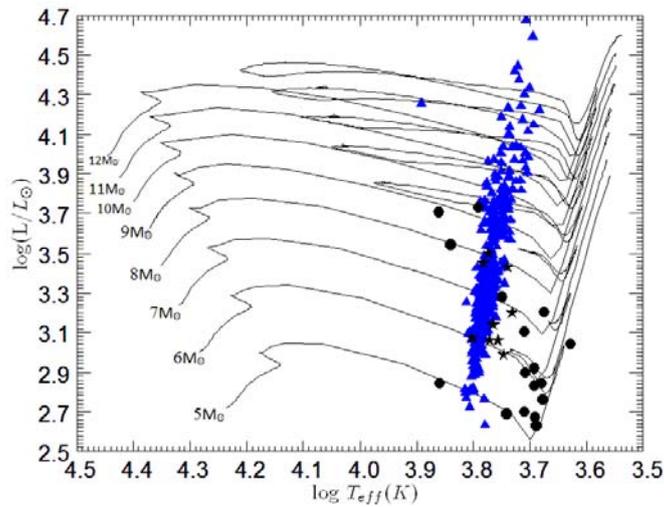

**FIGURE 2.** Same as FIGURE 1. The C-rate is used for the $^{14}N(p,\gamma)^{15}O$ reaction, see text. The triangles denote the observed Galactic Classical Cepheids by [7]. Observations by [10] are indicated by asterics and closed circles which represent Cepheid stars and yellow giants in open clusters respectively. Note the impressive reduction in the extension of the blue loops in the mass range $5M_\odot$ to $7M_\odot$ with this rate. This undesirable feature leaves a considerable number of observed stars unexplained.

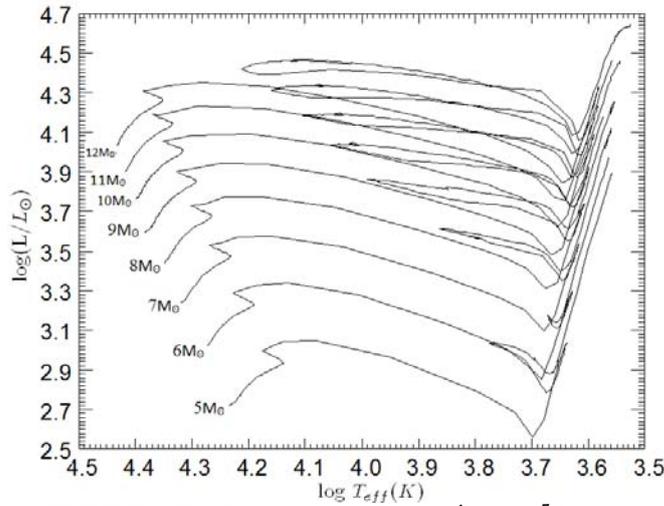

**FIGURE 3.** Same as FIGURE 1. The L-rate is used for the $^{14}N(p,\gamma)^{15}O$ reaction, see text. Also note the reduction in the extension of the blue loops in the mass range $5M_\odot$ to $7M_\odot$ with this rate.

We found that the $^{14}N(p,\gamma)^{15}O$ rate has a strong impact on the blue loops in the mass range comprising super-AGB stars, particularly $5M_\odot$ to $8M_\odot$. An appreciably reduced NACRE rate proposed on experimental ground can lead to a severe suppression of the blue loop. We have identified two effects of the $^{14}N(p,\gamma)^{15}O$ rate: one is the influence of this rate on the depth of the H-discontinuity during the first dredge up on the RGB prior to the loop phase, and the other effect is its impact on the efficiency of shell H-burning during this phase. These two effects are acting differently depending on the stellar mass range.

a. In case of stellar masses below $5M_\odot$, convection in the envelope creates a H-profile that is deep enough in mass, but the temperatures in the layers of shell H-burning are not high enough to re-activate the shell. The evolutionary track thus remains close to the RGB and no extended loop is formed. Thus, a modified N14 rate will not change this conclusion.

b. In case of stellar masses between $5M_\odot$ and $8M_\odot$, the situation is entirely different. The location of the H-discontinuity created by the envelope convection and the temperatures achieved in the region of the H-discontinuity are both crucial in determining the formation and extension of the loop. The effect of the N14 rate is most pronounced in this mass range and its efficiency plays a decisive role here because the temperatures in the proximity of the H-discontinuity are not as high as those in stars of masses above $8M_\odot$. Moreover, the less efficient shell H-burning with the weak C-rate causes an insufficient expansion of the envelope on the RGB, which results in a less deep H-profile. The combined effect of these factors causes a severely reduced loop.

c. In case of stellar masses above $8M_\odot$, the N14 rate has a marginal influence, because the shell source is strong enough owing to the high temperature encountered in the neighborhood of the H-discontinuity in these stars to initiate a well extended loop.

One possibility to restore the extension of blue loops is to invoke envelope overshooting, which places the H-discontinuity deeper in the star i.e. closer to the H-shell source. We have recalculated the evolution of the stars in the mass range $5M_\odot$ to $8M_\odot$ with the C-rate and L-rate, but with including exponential overshooting beyond the convective boundary as obtained with Schwarzschild criterion. We find that an overshooting distance of $0.2H_p$ is sufficient to restore the loop in the models that otherwise lack them.

## CONCULSION

Several works note the sensitivity of the blue loops to the N14 rate; however, its pronounced effect on a wide mass range is presented for the first time, which provides an insight on the behavior of the blue loop as a function of the stellar mass.

Finally, we emphasize that the N14 rate is a key reaction in this context and needs to be better constrained experimentally. In addition, if this rate would be less efficient by about a factor of two compared to the evaluation by NACRE, then, in this case the blue loops may be recovered only by introducing moderate envelope overshooting during the RGB. Indeed, stellar evolution conceals surprises which call for a careful treatment of all its conditions, and renders its astrophysical context rather challenging.

## ACKNOWLEDGMENTS

G.H. and M.E. thank the American University of Beirut (AUB) for supporting their research work. G.H. is thankful for the support she has received from the organiser of CSSP12 Prof. Livius Trache, as well as the URB at AUB that allowed her to present the current results in CSSP12. This work was also supported in part by the CNRSL under the grant number 111150-522276. She also thanks the IT Unit at the Faculty of Engineering & Architecture (FEA) and the Center of Advanced Mathematical Physics (CAMS) at AUB for using their computation facilities.